\documentclass[12pt]{article}
\usepackage{geometry} 
\usepackage{graphicx}

\parskip=\medskipamount
\def\caE{{\cal E}}

\def\al{\alpha}

\def\la{\lambda}  
\def\te{\theta}   

\def\k{\kappa}    
   
\def\IC{\relax{\rm l\kern-.50 em C}}
\def\IE{\relax{\rm l\kern-.16 em E}}
\def\IK{\relax{\rm l\kern-.18 em K}}
\def\IL{\relax{\rm I\kern-.18 em L}}
\def\IN{\relax{\rm I\kern-.18 em N}}
\def\IR{\relax{\rm I\kern-.18 em R}}

\font\tenfrak=eufm10  \font\sevenfrak=eufm7  \font\fivefrak=eufm5
\newfam\frakfam
\textfont\frakfam=\tenfrak
\scriptfont\frakfam=\sevenfrak
\scriptscriptfont\frakfam=\fivefrak

\def\wt{\widetilde}
\def\frac#1#2{{#1\over #2}}
\def\fracpd#1#2{\frac{\partial #1}{\partial #2}}

\def\Cos{\mathop{\rm C}\nolimits}    
\def\Sin{\mathop{\rm S}\nolimits}    
\def\Tan{\mathop{\rm T}\nolimits}    

\begin{document}

\title{ Curvature-dependent formalism, Schr\"odinger equation and energy levels for the harmonic oscillator on three-dimensional spherical and  hyperbolic spaces  
}
 
\author{
Jos\'e F. Cari\~nena$\dagger\,^{a)}$,
Manuel F. Ra\~nada$\dagger\,^{b)}$,
Mariano Santander$\ddagger\,^{c)}$ \\
$\dagger$
  {\sl Departamento de F\'{\i}sica Te\'orica and IUMA, Facultad de Ciencias} \\
  {\sl Universidad de Zaragoza, 50009 Zaragoza, Spain}  \\
$\ddagger$
  {\sl Departamento de F\'{\i}sica Te\'orica, Facultad de Ciencias} \\
  {\sl Universidad de Valladolid,  47011 Valladolid, Spain}
}

 \date{}
\maketitle

\begin{abstract}
 A nonlinear model representing the quantum harmonic oscillator
on the three-dimensional spherical and hyperbolic spaces,
$S_\k^3$ ($\kappa>0$) and $H_k^3$ ($\kappa<0$), is studied.  
The curvature  $\k$ is considered as a parameter and then 
the radial Schr\"odinger equation becomes a $\k$-dependent 
Gauss hypergeometric equation that can be considered as 
a $\k$-deformation of the confluent  hypergeometric equation 
that appears in the Euclidean case.
 The energy spectrum and the wavefunctions are exactly 
obtained in both the three-dimensional sphere $S_\k^3$ ($\kappa>0$)
and the hyperbolic space $H_k^3$ ($\kappa<0$).  
A comparative study between the spherical and the hyperbolic 
quantum results is presented.
\end{abstract}

\begin{quote}

{\sl Running title:}{\enskip}
The quantum harmonic oscillator on spaces with curvature.


\end{quote}
{\vfill}

\footnoterule
{\noindent\small
$^{a)}${\it E-mail address:} {jfc@unizar.es}  \\
$^{b)}${\it E-mail address:} {mfran@unizar.es} \\
$^{c)}${\it E-mail address:} {msn@fta.uva.es}
\newpage

\section{Introduction }

The study of  quantum problems in curved spherical spaces (positive constant curvature) was initiated by Schr\"odinger \cite{Sch40},  Infeld \cite{In41}, and Stevenson \cite{St41},    in t 1940 and 1941.   Infeld and Schild \cite{InSc45}  considered in 1945 a similar problem but in a
hyperbolic space (negative constant curvature). Later,  Barut {\sl et al} studied  a path integral treatment for the Hydrogen atom in a curved space 
of constant curvature, first in the spherical case \cite{BaInJ87} and then in the hyperbolic case  \cite{BaInJ90}. Since then other authors have studied similar problems on curved spaces with constant curvature making use of different approaches 
\cite{ComH85}-\cite{BaletalAnnPhys11}. 
Most of these papers are concerned with fundamental problems (previously studied at the classical level) but some authors have proved that this matter is also   important for  the study of certain questions related to  condensed matter  physcis as, for example,  the existence of Landau levels for the motion of a charged particle in a curved space \cite{Com87}-\cite{FaSh04} and, more recently, the study of quantum dots \cite{Gritsev01}-\cite{StoTu10}. 

It is clear that  spherical and hyperbolic spaces are endowed with quite different geometrical properties and this is the main reason why the studies of physical systems on spherical and hyperbolic spaces are usually carried out in a separated way (see, e.g., most of the above mentioned references) since their physical properties turn out to be also different. In spite of this, it has been proved that certain problems (in fact, those related with superintegrable potentials) can be studied making use of a joint approach valid for the two types of spaces. 

The paper is concerned with the study of the quantum harmonic oscillator 
on  three-dimensional spherical and hyperbolic spaces making use of a set of coordinates $(r,\te,\phi)$ obtained by introducing a small change in the radial part of the geodesic spherical  coordinates.   
It can be considered as a new paper in a series devoted to the study  of classical  
\cite{RaSa02I}-\cite{CRS08Jmp} and quantum \cite{CRS04Rmp} -\cite{CRS11Jmp} systems on Riemannian conÞguration spaces with constant  curvature $\k\ne 0$.  
We follow an approach that can be  summarized in the following two points:

\begin{enumerate}
\item[ (i)]  All the mathematical expressions will depend of the curvature $\k$ as a parameter.  So, the first step is to obtain general $\k$-dependent properties.  Then, 
 the second step is to particularize for the values  $\k>0$, $\k=0$, or $\k<0$, and obtaining, in such a way, the corresponding property for the physical system  on the sphere $S_{\k}^3$, on the Euclidean space $\IE^3$,  or on the hyperbolic space $H_{\k}^3$, respectively. 

\item[ (ii)] The idea is to formulate the results in explicit dependence of the curvature $\k$ and to study the changes of the dynamics when $\k$ varies. 
\end{enumerate}

We mention now two  points that are important for the study presented in this paper. The first one  is related to the geometric approach and the other to the dynamis.

\begin{itemize}

\item    The differential element of distance $ds_{\k}$,  in the family $M_{\k}^3 =
(S_{\k}^3, \IE^3, H_{\k}^3)$  of three-dimensional spaces with constant
 curvature $\k$, can be written is some different but equivalent ways 
 (this question is discussed in \cite{CRS11IntJ}--\cite{CRS11Jmp}; see also \cite{BaVo86}). 
For example, if we make use of the following $\kappa$-dependent trigonometric
(either circular, parabolic or hyperbolic) functions
$$
 \Cos_{\k}(x) = \cases{
  \cos{\sqrt{\k}\,x}       &if $\k>0$, \cr
  {\quad}  1               &if $\k=0$, \cr
  \cosh\!{\sqrt{-\k}\,x}   &if $\k<0$, \cr}{\qquad}
  \Sin_{\k}(x) = \cases{
  \frac{1}{\sqrt{\k}} \sin{\sqrt{\k}\,x}     &if $\k>0$, \cr
  {\quad}   x                                &if $\k=0$, \cr
  \frac{1}{\sqrt{-\k}}\sinh\!{\sqrt{-\k}\,x} &if $\k<0$, \cr}
$$
then it can be written as follows in geodesic spherical  coordinates $(\rho,\te,\phi)$ 
\begin{equation}
 ds_{\k}^2 = d\rho^2 + \Sin_\k^2(\rho)\,(d\te^2 + \sin^2 \te\,d\phi^2) \,,
\end{equation}
(note that this formalism is intrisic and $\rho$ denotes the distance along a geodesic on the manifold $M_{\k}^3$ and not the radius of a sphere).  Nevertheless in the following we will use a new radial variable $r$ given by $r=\Sin_\k(\rho)$ so the expression of $ds_{\k}^2$ in the coordinates $(r,\te,\phi)$ becomes 
\begin{equation}
 ds_{\k}^2 = \frac{dr^2}{1 - \k \, r^2} + r^2\,d\te^2  + r^2\sin^2 \te\,d\phi^2  \,,
\end{equation}
so it reduces to
\begin{eqnarray*}
 ds_1^2  &=&   \frac{dr^2}{1 -\, r^2} + r^2\,(d\te^2 + \sin^2 \te\,d\phi^2) \,, \cr
 ds_0^2  &=&   dr^2  + r^2\,(d\te^2 + \sin^2 \te\,d\phi^2)    \,, \cr
 ds_{-1}^2 &=&  \frac{dr^2}{1 +\, r^2} + r^2\,(d\te^2 + \sin^2 \te\,d\phi^2) \,,
\end{eqnarray*}
in the three particular cases of the unit sphere, the Euclidean plane,  and the `unit` 
Lobachewski plane.

\item   
The harmonic oscillator in the space of constant curvature $\k$ has a potential $U_\k$   which, when written in the system $(\rho,\te,\phi)$,  is given by 
\begin{equation}
 U_\k(\rho) = \frac{1}{2}\,\al^2 \, \Tan_{\k}^2(\rho)  \,, 
\end{equation}
where $\Tan_\k(\rho)$ denotes the $\k$-dependent tangent. It is `central' in the  sense that it depends only on the geodesic distance $\rho$ to a fixed center in $M_{\k}^3$.  When the curvature $\k$ is positive then  the potential tends to infinity at the sphere `equator' (with the north pole placed in the center of forces), which corresponds to a finite value $\rho=\pi/(2\sqrt{\k})$; the harmonic oscillator on the $\k>0$ sphere splits the configuration space into two halves with an infinite potential wall on the equator; so the spherical harmonic oscillator motion is confined to just one of these halves. As stated above, we will use the  coordinates $(r,\te,\phi)$ wherein the potential $U_\k$ becomes  
\begin{equation}
  U_\k(r) = \frac{1}{2}\,\al^2 \,\Bigl(\frac{r^2}{1 - \k\,r^2} \Bigr) \,. 
\end{equation}
Of course, for $\k>0$, while the geodesic radial coordinate $\rho$ in the range $[0, \pi/\sqrt{\k}]$ allows covering the whole sphere (with coordinate singularities at both ends of the range), the range $[0, 1/\sqrt{\k}]$ of the radial coordinate $r$  covers naturally only the upper half of the sphere; this matches perfectly with the nature of the harmonic potential for the positive curvature case, which has a infinite wall at the  boundary of the domain naturally covered by the coordinate $r$.  

 Figure 1 shows that the standard  Euclidean potential ($\k=0$) represents a borderline between two different behaviors. If $\k>0$, then the potential tends to infinity when  $r^2\to 1/\k$. Then, in the case $\k<0$  the potential is well defined for all the values of $r$ and it is even bounded when $r\to\infty$. 

The potential $U_\k(r)$ is interpreted as describing the harmonic oscillator in the spaces $M_{\k}^3 = (S_{\k}^3, \IE^3, H_{\k}^3)$ because of two reasons: First, it fulfills the Euclidean limit in the sense that when $\k\to 0$ it becomes the well known potential of the isotropic harmonic oscillator in the Euclidean space (this is a necessary condition).  Secondly, and even more important, this potential is singled out amongst other possibilities with the same Euclidean limit by the condition of being  superintegrable (some details are provided in the next section). 

\end{itemize}

  The plan of the article is as follows: In Sec. 2, we obtain the expression of the $\k$-dependent quantum Hamiltonian $H(\k)$. In fact this section is mainly related to a previous study presented in Ref. \cite{CRS11IntJ}.  In Sec. 3,   the $\k$-dependent
   Schr\"odinger equation is solved and then the properties of the spherical $\k>0$ and the hyperbolic $\k<0$ cases are studied with detail. 
Finally, in Sec. 4, we make some final comments.

\section{$\k$-dependent quantum Hamiltonian  }

The construction of the classical $\kappa$-dependent system and the transition from the classical $\kappa$-dependent system to the quantum one was studied in \cite{CRS11IntJ}.  The main idea is to follow a method used in some previous references as \cite{CRS07AnPh2}--\cite{CRS11Jmp} that considers  the quantization of the Noether momenta as a first step; in this way the Hamiltonian $H(\k)$ is obtained  as a self-adjoint operator with respect to an appropriate $\k$-dependent measure.

 The Lagrangian $L$ of the geodesic motion ($\k$-dependent kinetic term $T(\k)$ without a potential) on the three-dimensional spaces $(S_{\k}^3, \IE^3, H_{\k}^3)$  is given by 
\begin{equation}
 L = T(\k) = (\frac{1}{2})\,\Bigl(\frac{v_r^2}{1 - \k\,r^2} + r^2\,v_{\te}^2
 + r^2\sin^2 \te\,v_{\phi}^2\Bigr)    \,, 
\end{equation}
where the parameter $\k$ can be positive (spherical case), null (Euclidean space) and negative (hyperbolic space). In the spherical case,  the study of the dynamics is restricted to the interior of the interval $r^2<1/\k$ where the kinetic energy is a positive-definite function. As a consequence of the six-dimensional geometric symmetry of this sytem, encompassed by a group isomorphic to either $SO(4), ISO(3), SO(1,3)$ according to $\k>, =, < 0$. this Lagrangian possesses a total of six Noether symmetries. Three of them are related to the common rotational $SO(3)$ symmetry present in all spaces, and are other are specific to each space.

The latter are the three $\k$-dependent Noether symmetries 
\begin{eqnarray*}
  X_{1}(\k) &=& \sqrt{1 - \k \, r^2} \,\Bigl[(\sin{\te}\cos{\phi})\,\fracpd{}{r}
  + \frac{1}{r} \bigl[(\cos{\te}\cos{\phi})\,\fracpd{}{\te}
  - (\frac{\sin{\phi}}{\sin\te})\,\fracpd{}{\phi}  \bigr] \Bigr]  \,,\cr
  X_{2}(\k) &=& \sqrt{1 - \k \, r^2} \,\Bigl[ (\sin{\te}\sin{\phi})\,\fracpd{}{r}
  + \frac{1}{r} \bigl[(\cos{\te}\sin{\phi})\,\fracpd{}{\te}
  + (\frac{\cos{\phi}}{\sin\te})\,\fracpd{}{\phi} \bigr] \Bigr]\,,\cr
  X_{3}(\k) &=& \sqrt{1 - \k \, r^2} \,\Bigl[\,(\cos{\te})\,\fracpd{}{r}
  - \frac{1}{r} \sin{\te}\,\fracpd{}{\te}   \Bigr]\,,
\end{eqnarray*}
with associated constants of motion 
\begin{eqnarray*}
 P_1(\k) &=& (\sin{\te}\cos\phi)\,\frac{v_r}{\sqrt{1 - \k\,r^2}} + (r\sqrt{1 - \k\,r^2})
   \bigl[(\cos{\te}\cos{\phi})\,v_\te - (\sin{\te}\sin{\phi})\,v_\phi\bigr]\,,\cr
 P_2(\k) &=& (\sin{\te}\sin\phi)\,\frac{v_r}{\sqrt{1 - \k\,r^2}} + (r\sqrt{1- \k\,r^2})
   \bigl[(\cos{\te}\sin{\phi})\,v_\te + (\sin{\te}\cos{\phi})\,v_\phi\bigr] \,,\cr
 P_3(\k) &=& (\cos{\te})\,\frac{v_r}{\sqrt{1 - \k\,r^2}} -  (r\sqrt{1 - \k\,r^2})\sin{\te}\,v_\te \,.
\end{eqnarray*}
while the former are  the three $\k$-independent Noether symmetries that coincide with the Euclidean symmetries corresponding to the space isotropy  
 $$
  Y_1 =   -\sin\phi\,\fracpd{}{\te}
      - (\frac{\cos{\phi}}{\tan \te})\,\fracpd{}{\phi} \,,{\qquad}
  Y_2 =   \cos\phi\,\fracpd{}{\te}
      - (\frac{\sin{\phi}}{\tan \te})\,\fracpd{}{\phi} \,,{\qquad}
  Y_3 =   \fracpd{}{\phi} \,,
$$
leading to the three ($\k$-independent) components of the angular momentum 
\begin{eqnarray*}
   J_1  &=&-\,r^2(\sin\phi\,v_\te + \sin\te\cos\te\cos\phi\,v_\phi) \,,\cr
   J_2  &=&  r^2 (\cos\phi\,v_\te - \sin\te\cos\te\sin\phi\,v_\phi) \,,\cr
   J_3  &=&  r^2\sin^2{\te}\,v_\phi\,.
\end{eqnarray*}

The $\k$-dependent Hamiltonian representing the harmonic oscillator on the spaces $(S_{\k}^3, \IE^3, H_{\k}^3)$  is given by 
\begin{equation}
 H(\k) = \bigl(\frac{1}{2}\bigr)\,\Bigl[\, (1 - \k\,r^2)\,p_r^2
 + \frac{1}{r^2}\,(p_{\te}^2 + \frac{p_{\phi}^2}{\sin^2\te\,})\,\Bigr]
 + \frac{1}{2}\,(m\al^2)\Bigl(\frac{r^2}{1 - \k\,r^2} \Bigr)  \,, 
\label{Hclas}
\end{equation}
and it can also be written  as follows 
\begin{equation}
 H(\k) = \bigl(\frac{1}{2 m}\bigr)\,
 \Bigl[ P_1^2 + P_2^2 +P_3^2 + \k\,(J_x^2 +J_y^2+J_z^2)\Bigr]
 + \frac{1}{2}\,(m\al^2)\Bigl(\frac{r^2}{1 - \k\,r^2} \Bigr)  \,,
\end{equation}
where $P_j$ and $J_j$, $j=1,2,3$, denote now the Hamiltonian versions of the corresponding Noether momenta  obtained above in the Lagrangian notation. Here, it is worth mention that the free part of the Hamiltonian is proportional to the quadratic Casimir operator in the Lie algebra of the geometric symmetries, reducing to the square of the linear momentum only in the case $\k=0$, but including also a term with the square of the angular momentum otherwise.

The quantum Hamiltonian $\widehat{H}(\k)$ must be an operator obtained from $H(\k)$ that must be self-adjoint in the space $L_\k^2(d\mu_\k)$  where $d\mu_\k$ denotes the measure $d\mu_\k$ 
$$
  d\mu_\k =  \Bigl(\frac{r^2\,\sin\te}{\sqrt{1-\k\,r^2}}\Bigr)\,dr\,d\te\,d\phi  
$$
and the particular form of the Hilbert space $L_\k^2(d\mu_\k)$ depends on $\k$ as follows 
\begin{itemize}
\item[(i)]  In the hyperbolic $\k<0$ case, the space $L_\k^2(d\mu_\k)$ can be identified with $L^2(\IR^3,d\mu_\k)$. 
\item[(ii)] In the spherical $\k>0$ case, the space $L_\k^2(d\mu_\k)$  can be identified with $L_0^2(I_\k\times\IR^2,d\mu_\k)$ where $I_\k$ denotes the interval $[0,1/\sqrt{\k}]$ and the subscript means that the functions must vanish at the  endpoints $r=0$ and $r=1/\sqrt{\k}$. 
\end{itemize}
The first step is to obtain the expressions of the operators 
$\widehat{P_1}$,  $\widehat{P_2}$,  and $\widehat{P_3}$, 
representing the quantum version of of the Noether  momenta 
$P_1$, $P_2$, an $P_3$, as self-adjoint operators 
 in the space $L_\k^2$. They are given by 
\begin{eqnarray*}
    \widehat{P_1}&=&  -\,i\,\hbar\,
 \sqrt{1 - \k \, r^2} \,\Bigl[ (\sin{\te}\cos{\phi})\,\fracpd{}{r}
  + \frac{1}{r} \bigl[(\cos{\te}\cos{\phi})\,\fracpd{}{\te}
  - (\frac{\sin{\phi}}{\sin\te})\,\fracpd{}{\phi}  \bigr] \Bigr] \,,\cr
    \widehat{P_2}&=& -\,i\,\hbar\,
\sqrt{1 - \k \, r^2} \,\Bigl[ (\sin{\te}\sin{\phi})\,\fracpd{}{r}
  + \frac{1}{r} \bigl[(\cos{\te}\sin{\phi})\,\fracpd{}{\te}
  + (\frac{\cos{\phi}}{\sin\te})\,\fracpd{}{\phi} \bigr] \Bigr]   \,,\cr
    \widehat{P_3}&=& -\,i\,\hbar\,
 \sqrt{1 - \k \, r^2} \,\Bigl[\,(\cos{\te})\,\fracpd{}{r}
  - \frac{1}{r} \sin{\te}\,\fracpd{}{\te} \Bigr]  \,. 
\end{eqnarray*}
The quantum operators $\widehat{J_i}$, $i=1,2,3$, are $\k$-independent and therefore they coincide with the Euclidean ones 
$$
    \widehat{J_1}=  \,i\,\hbar\,\Bigl[\,\sin\phi\,\fracpd{}{\te}
      + (\frac{\cos{\phi}}{\tan \te})\,\fracpd{}{\phi}\,\Bigr]\,,\quad
    \widehat{J_2} =  -\,i\,\hbar\,\Bigl[\,\cos\phi\,\fracpd{}{\te}
      - (\frac{\sin{\phi}}{\tan \te})\,\fracpd{}{\phi}\,\Bigr]   \,,\quad
    \widehat{J_3} =   -\,i\,\hbar\, \fracpd{}{\phi} \,. 
$$
Then we have that the quantum  Hamiltonian   $\widehat{H}(\k)$ that is given by 
\begin{equation}
 \widehat{H}(\k) =\bigl(\frac{1}{2 m}\bigr)
 \Bigl[ \widehat{P_1}^2 + \widehat{P_2}^2 +\widehat{P_3}^2
 + \k\,(\widehat{J_1}^2 +\widehat{J_2}^2+\widehat{J_3}^2)\Bigr]
  + \frac{1}{2}\,(m\al^2)\Bigl(\frac{r^2}{1 - \k\,r^2} \Bigr)  \,,
\label{Hq1}
\end{equation}
is represented by the following differential operator
\begin{equation}
 \widehat{H} = - \frac{\hbar^2}{2 m} \Bigl[(1 - \k\,r^2)\,\fracpd{^2}{r^2}
 + \frac{2 - 3\k\,r^2}{r}\,\fracpd{}{r} + \frac{1}{r^2} \Bigl(
 \fracpd{^2}{\te^2}+\frac{1}{\sin^2\te}\fracpd{^2}{\phi^2}+\frac{1}{\tan\te}\fracpd{}{\te}\Bigr) \Bigr]
 + \frac{1}{2}(m\al^2)\Bigl(\frac{r^2}{1 - \k\,r^2} \Bigr)    \,, 
\label{Hq2}
\end{equation}
that is self-adjoint with respect the measure $d\mu_\k$ and it  satisfies the appropriate Euclidean limit
$$
 \lim_{\k\to 0} \widehat{H}(\k) =   - \frac{\hbar^2}{2 m}\,
 \Bigl[\,\fracpd{^2}{r^2} + \frac{2}{r}\,\fracpd{}{r} + \frac{1}{r^2}\,
 \Bigl( \fracpd{^2}{\te^2}+\frac{1}{\sin^2\te}\fracpd{^2}{\phi^2}
 +\frac{1}{\tan\te}\fracpd{}{\te}\Bigr) \Bigr]   + \frac{1}{2}\,(m\al^2) r^2\,.
$$

We close this section with the following remarks and observations.

\begin{enumerate}  

\item 
The requirement to have the correct Euclidean limit leaves a lot of possibilities open for a `harmonic oscillator  potential in the curved space'. But the potential chosen is singled out if we require also that the curved potential be also superintegrable. Indeed, for this potential $U_\k(r)$, there is a full set of constants of motion given by the $\k$-dependent functions $F_{ij}(\k)$ defined by 
$$
 F_{ij}(\k) = P_i P_j + \al^2 X_i X_j \,,{\quad} i,j=1,2,3,  
$$
with $X_i$, $i=1,2,3$,  given by 
$$
 X_1 = \frac{r\,\sin\te \cos\phi }{\sqrt{1 - \k\,r^2}} \,,\quad 
 X_2 = \frac{r\,\sin\te \sin\phi}{\sqrt{1 - \k\,r^2}}  \,,\quad 
 X_3 = \frac{r\,\cos\te }{\sqrt{1 - \k\,r^2}} \,.  
$$
Of course $F(\k)$ with components $F_{ij}(\k)$, $i,j=1,2,3$,  represents the curved version of the Fradkin tensor \cite{Fr65}.  Notice that the expression of $F(\k)$ depends of the Noether momenta instead of the canonical momenta. 

\item  The measure $d\mu_\k$, that was obtained as the unique measure (up to a multiplicative constant) invariant under the Killing vectors \cite{CRS11IntJ}, coincides with the corresponding  Riemann volume in a space with curvature $\k$. 

\item  The (free part of the) quantum Hamiltonian we have obtained $\widehat{H}(\k)$ turns out to coincide with the one obtained by making use of the Laplace-Beltrami operator for the space under consideration. This is no surprise, of course. While the end result is the same, we want to emphasize that the logic in the argument is somewhat different to the usual because  Laplace-Beltrami  quantization procedure leads directly to the expression of the quantum Hamiltonian  without a previous quantization of the momenta. The standard procedure in the Euclidean case is to first quantize the momenta (i.e. to identify them as self-adjoint operators) and then to obtain the quantum version of the Hamiltonian. We have translated this momentum-approach to the spaces with curvature $\k$ but changing the quantization of the canonical momenta by the quantization of the Noether momenta which are taken as the basic objects. 

\item  One additional reason for the quantization via the Noether momenta is that it also seems  appropriate for the quantization of systems with a position dependent mass (PDM). In fact, there is  a great interest in the study of the quantization of systems with a PDM not only for the applications to condensed matter physics, but also because there is an important problem at the starting level of quantization; since if the mass $m$ becomes a spatial function,  then the quantum version of the mass no longer commutes with the momentum. 
A Hamiltonian system in a space with curvature $\k$ can also be considered as a very particular PDM system (in this case de the mass $m$ is not an effective mass but it becomes a spatial function as a consequence of the geometry). We think that the quantization, as a first step, of the Noether momenta is an appropriate method for the quantization of the PDM systems. 

\end{enumerate}

\section{$\k$-dependent Schr\"odinger equation and wavefunctions}


The Schr\"odinger equation $\widehat{H}(\k)\,\Psi = E\,\Psi$ 
leads to the following $\k$-dependent differential equation
\begin{eqnarray}
 \Biggl[\,- \frac{\hbar^2}{2 m}\, \Bigl[(1 - \k\,r^2)\,\fracpd{^2}{r^2}
 + \frac{2 - 3\k\,r^2}{r}\,\fracpd{}{r} &+& \frac{1}{r^2}\,\Bigl(
 \fracpd{^2}{\te^2}+\frac{1}{\sin^2\te}\fracpd{^2}{\phi^2}+\frac{1}{\tan\te}\fracpd{}{\te}\Bigr) \Bigr]      \cr
  &+& \frac{1}{2}\,(m\al^2)\,\Bigl(\frac{r^2}{1 - \k\,r^2} \Bigr)\,\Biggr]\,\Psi =
  E\,\Psi   \,. 
\end{eqnarray}
Thus, as $U_\k(r)$ is a central potential for all the values of $\k$,  we can assume that $\Psi(r,\te,\phi)$ can be factorized of the form
$$
 \Psi(r,\te,\phi) = R(r) \,Y_{lm}(\te,\phi) \,,
$$
where $R$ is a function of $r$ and $Y_{lm}(\te,\phi)$ are the
standard $\k$-independent spherical harmonics
 $$
 \Bigl( \fracpd{^2}{\te^2} +\frac{1}{\sin^2\te}\fracpd{^2}{\phi^2}
 +\frac{1}{\tan\te}\fracpd{}{\te} \Bigr)\,Y_{lm} = -\,l(l+1)\,Y_{lm}.
$$
Then we arrive at the following $\k$-dependent radial equation
$$
  \Biggl[\,- \frac{\hbar^2}{2 m}\, \Bigl[(1 - \k\,r^2)\,\frac{d^2}{dr^2}
  + \frac{2 - 3\k\,r^2}{r}\,\frac{d}{dr} - \frac{l(l+1)}{r^2}\Bigr]
  + \frac{1}{2}\,(m\al^2)\,\Bigl(\frac{r^2}{1 - \k\,r^2} \Bigr)\,\Biggr]\,R= E\,R
  \,,\quad R = R(r) \,. 
$$

 It is  convenient to change the parameter $\al^2$ in the potential
to the form $\al^2\,\to\, \al^2 - (\k\,\hbar/m)\al$ 
(this change is done by similarity with the result obtained when using the Schr\"odinger factorization method for the one-dimensional non-linear oscilator studied in \cite{CRS04Rmp}, \cite{CRS07AnPh2}) and  introduce  
dimensionless variables $(\rho,\wt{\k},\caE)$ defined by  
$$
  r = \Bigl(\sqrt{\frac{\hbar}{m\al}}\,\Bigr)\,\rho \,,{\quad}
 \k = \Bigl(\frac{m\,\al}{\hbar}\Bigl)\,\wt{\k}  \,,{\quad}
  E = (\hbar\al)\,\caE  \,,{\quad}\k\,r^2 = \wt{\k} \,\rho^2\,, 
$$
so that we obtain 
\begin{equation}
  \rho^2\,(1 - \wt{\k}\,\rho^2)\,R''  +\rho\,(2 - 3\wt{\k}\,\rho^2)\,R'
  - (1 - \wt{\k})\,\Bigl(\frac{\rho^4}{1 - \wt{\k}\,\rho^2} \Bigr)\,R
+ \Bigl[2\,\caE\,\rho^2 - l(l+1) \Bigr]\,R   = 0   \,,
\end{equation} 
that represents a $\k$-dependent deformation of the Euclidean
differential equation
$$
 R'' + \frac{2}{\rho}\, R' - \rho^2\,R + \Bigl[2\,\caE -
\frac{l(l+1)}{\rho^2} \Bigr]\,R = 0   \,.
$$
We assume the following factorization for the function $R$:
$$
 R = f(\rho,\wt{\k})\,(1 - \wt{\k}\,\rho^2)^{\,(1/2\wt{\k})} \,,
$$
so that 
$$
 \lim{}_{\wt{\k}{\to}0}\,R(\rho,\wt{\k}) = f(\rho)\,e^{-\,(1/2)\,\rho^2}\,.
$$
Then the function $f(\rho)$  must be solution of 
\begin{equation}
  \rho^2\,(1 - \wt{\k}\,\rho^2)\,f'' +\rho\,(2 - 2 \rho^2 - 3\wt{\k}\,\rho^2)\,f'
 + \Bigl[(2\,\caE - 3)\,\rho^2 - l(l+1) \Bigr]\,f = 0   \,. 
\end{equation} 
This equation can be solved by using the method of Frobenius.
The solution near the regular singular point $\rho=0$ can be written as follows:
$$
  f = \rho^\mu\,g(\rho,\wt{\k}) \,, 
$$
where $\mu$ is a solution of the indicial equation and $g$ is an analytical function with a $\k$-dependent power series
$$
 g  = \sum_{n=0}^{\infty}\,  g_n \rho^n
 = g_0 + g_1 \rho + g_2 \rho^2 + g_3 \rho^3 + \dots \quad (g_0\ne 0) \,. 
$$
Then it is proved that $\mu$ must take one of the two values
$\mu_1 = l$ or $\mu_2=-l-1$.  Considering $\mu=l$, in order to have
$R$  well defined at the origin, we arrive at
\begin{equation}
  \rho\,(1 - \wt{\k}\,\rho^2)\,g'' +
  \Bigl[2(l + 1) - (2 + 3 \wt{\k} + 2 \wt{\k}\,l)\rho^2\Bigr]\,g'
 + \Bigl[(2\,\caE - 3) - (2 + 2\wt{\k} + \wt{\k}\,l)\,l \Bigr]\,\rho\,g  = 0   \,,
\end{equation}  
and then the $\k$-dependent recursion relation leads to the vanishing
of all the odd coefficients, $ g_1 = g_3 = g_5 = g_7 = \dots = 0$,
so that it is a series with only even powers of $\rho$ and a
radius of convergence $R_c$ given by $R_c =
1/\sqrt{\,|\,\wt{\k}\,|\,}$ (determined by the presence of the
second singularity). The even powers dependence suggests to
introduce the new variable $z=\rho^2$ so that the equation becomes
\begin{equation}
 z\,(1 - \wt{\k}\,z)\,g_{zz}'' + \frac{1}{2} \Bigl[(2 l + 3) - 2 (1 + 2\wt{\k} + \wt{\k}\, l)z \Bigr]\,g_{z}' + \frac{1}{4}\Bigl[(2\,\caE - 3) - (2 + 2\wt{\k} + \wt{\k}\,l)\,l \Bigr]\,g  = 0   \,,
\label{Ecg(zk14)}
\end{equation}
In the Euclidean case this equation reduces to 
$$
 z\,g_{zz}'' +  \Bigl[(l + \frac{3}{2}) - z \Bigr]\,g_{z}' - \frac{1}{2}\Bigl[(l + \frac{3}{2}) - \caE \Bigr]\,g  = 0   \,,
$$
whose solution regular at $z=0$ is a confluent hypergeometric function  
$$
  g(\rho) = {}_1F_1(a\,;c\,;\rho^2)\,,{\quad} 
  a = \frac{1}{2}\Bigl[(l + \frac{3}{2}) - \caE \Bigr]\,,{\quad} 
  c = l + \frac{3}{2}  \,. 
$$
The boundary conditions at $\rho=0$ and $\rho=\infty$ (Sturm-Liouville problem) leads to the associated Laguerre polynomials.

In the general non-Euclidean $\wt{\k}\ne 0$ case it is convenient
to introduce the change $t=\wt{\k}\, z$. Then, equation
(\ref{Ecg(zk14)}) reduces to
\begin{equation} 
 t\,(1 - t)\,g_{tt}'' + \Bigl[(l+\frac{3}{2}) - \frac{1}{\wt{\k}} 
 (1 + 2\wt{\k} + \wt{\k} l)\,t\Bigr]\,g_t'  + \frac{1}{4\wt{\k}}
 \Bigl[(2\,\caE - 3 - 2 l) - \wt{\k}\,l\,(l+2)\, \Bigr]\,g  = 0   \,,
\label{Ecg(tk15)}
\end{equation}
that is a Gauss hypergeometric equation
$$
 t\,(1 - t)\,g_{tt}'' + [c-(1+a_\k+b_\k)t]\,g_t' - a_\k b_\k g = 0 \,,
$$
with
$$
 c = l+\frac{3}{2}    \,,\qquad
 a_\k + b_\k = \frac{1}{\wt{\k}} + l + 1 \,,\qquad
 a_\k  b_\k = -\,\frac{1}{4\wt{\k}}\Bigl[(2\,\caE - 3 - 2 l) - \wt{\k}\,l\,(l+2) \Bigr]   \,,
$$
and the solution regular at $t=0$ is the hypergeometric function
$$
 g(t,\wt{\k}) = {}_2 F_1(a_\k,b_\k;c\,; t) \,,\qquad
 {}_2 F_1(a_\k,b_\k;c\,; t) = 1 + \sum_{n=1}^{\infty}\,
 \frac{(a_\k)_n\,(b_\k)_n}{(c)_n} \,\frac{t^n}{n\,!} \,,
$$
with $a_\k$ and $b_\k$ given by
$$
 a_\k =  \frac{1}{2\,\k}\bigl(\,A_\k \pm \sqrt{B_\k}\,\bigr)\,,\
 b_\k =  \frac{1}{2\,\k}\bigl(\,A_\k \mp \sqrt{B_\k}\,\bigr)
$$
(indeed, in wiew of the symmetry, there is no real restriction if one takes only the upper sign in both relations), where  
$$
 A_\k = 1 + \wt{\k} (l+1)  \,,\quad
 B_\k = 1 + (2\caE - 1)\wt{\k} + \wt{\k}^2   \,.
$$
The equation (\ref{Ecg(tk15)}) has a singularity,  when $\wt{\k}>0$,  at $t=1$ that correspond to
$z=1/\wt{\k}$ (or $r=1/\sqrt{\wt{\k}}$).  If the origin $r=0$ is
placed in the north pole of the sphere then this singularity is
just placed at the equator.  The property of regularity of the
solutions leads to analyze the existence of particular solutions
well defined at this point.  The polynomial solutions  appear when
one of the two $\k$-dependent coefficients, $a_\k$ or $b_\k$,
coincide with zero or with a negative integer number:
$$
 a_\k = -\,n_r\,,\quad {\rm or}\quad
 b_\k = -\,n_r\,,\qquad  n_r=0,1,2,\dots
$$
Then, in this case,  the coefficient $\caE$, that represents the
energy, is restricted to one of the following values:
$$
  \caE_{n_r,l} = (2n_r + l + \frac{3}{2}) + \frac{1}{2}\,\wt{\k}\,(2n_r + l)(2n_r + l+2) \,,
$$
and the hypergeometric series ${}_2 F_1(a_\k,b_\k,c\,; \wt{\k} z)$
reduces to a polynomial of degree $n_r$.

The differential equation 
$$
 a_0  g'' + a_1 g' + \la \rho\, g = 0  \,,
$$
with
$$
 a_0 =  \rho\,(1 - \wt{\k}\,\rho^2) \,,{\quad}  
 a_1 =  [2(l + 1) - (2 + 3 \wt{\k} + 2 \wt{\k}\,l)\rho^2   \,,{\quad}
 \la = (2\,\caE - 3) - (2 + 2\wt{\k} + \wt{\k}\,l)\,l  \,,
$$
together with the  boundary conditions in the points $\rho_1=0$ and $\rho_2=\rho_\k$, determine a singular Sturm-Liouville problem  that is formally self-adjoint and if 
 the  boundary conditions are appropriately defined  then the operator is symmetric. Then the eigenfunctions corresponding to distinct eigenvalues are orthogonal with respect to the weight function $q= \rho^{2(l+1)}(1 - \wt{\k}\,\rho^2)^{1/\wt{\k}-1/2}$; note that this involves the value of the curvature. More concretely we have  
\begin{itemize}
\item[(i)]  In the spherical $\k>0$ case, the function $g$ must vanish in the point $\rho_2=\rho_\k$ and the eigenfunctions are orthogonal in the interval $[0,\rho_\k]$ with $\rho_\k=1/\sqrt{\wt{\k}}$. 
\item[(ii)]  In the hyperbolic $\k<0$ case,  the function $g$ must satisfy the property $g\to 0$ when $\rho\to\infty$ and the eigenfunctions are orthogonal in the interval $[0,\infty)$. 
\end{itemize}

This statement is just a consequence of the properties of the Sturm-Liouville problems.   The $\k$-dependent differential equation for the function $g(\rho)$ is not  self-adjoint since $a'_0 \ne a_1$ but it can be reduced to self-adjoint form by making  use of the following integrating factor
$$
  \mu = \bigl(\frac{1}{a_0}\bigr)\,e^{{\int}(a_1/a_0)\,dr}
 =   \rho^{2l+1}(1 - \wt{\k}\,\rho^2)^{1/\wt{\k}-1/2}  \,,
$$
so that the equation becomes
\begin{equation}
  \frac{d}{d\rho}\Bigl[\,p(\rho,\wt{\k})\,\frac{dg}{d\rho}\,\Bigr] +  \la\,  q(\rho,\wt{\k})\,g = 0 \,,
 {\quad}   {\rm \la\  is\ a\ constant} \,,
 \label{EcSLg(ro)} \end{equation}
where $p(\rho,\wt{\k})=  \mu\,a_0$ and $q(\rho,\wt{\k})$ is given by 
$$
  q(\rho,\wt{\k}) = \rho^{2(l+1)} (1 - \wt{\k}\,\rho^2)^{1/\wt{\k}-1/2}   
  =\Bigl(\rho^{2l } (1 - \wt{\k}\,\rho^2)^{1/\wt{\k}} \,\Bigr)\Bigl(\frac{\rho^2}{\sqrt{1-\wt{\k}\,\rho^2}}\Bigr)  \,.
$$
Note that this problem is  singular  in the two cases but in a different way: (i) If $\k$ is positive because the function $p(\rho,\wt{\k})$ vanish at the  boundary point $\rho_2=\rho_\k$;  (ii) If $\k$ is negative, then the problem is also singular since it is defined in the semi-infinite positive real line $\IR^+$. Nevertheless,   the properties of the  Sturm-Liouville problems state that even in
these cases the eigenfunctions of the problem are orthogonal with
respect the function $q(\rho,\wt{\k})$.

To sum up, the essential result we have obtained is the following: for either value of the curvature, the radial wavefunction $R(\rho)$ which can appear together the usual spherical harmonic $Y_{lm}(\te,\phi)$ and which is regular at $\rho=0$ is (a multiple of)
\begin{equation}
r^l(1 - k\, r^2)^{\,(1/2\wt{\k})}
{}_2 F_1(\frac{1}{2\,\k}\bigl(\,A_\k \pm \sqrt{B_\k}\,\bigr),
\frac{1}{2\,\k}\bigl(\,A_\k \mp \sqrt{B_\k}\,\bigr),l+3/2\,; \,\k\, r^2) \,.
\end{equation}

The associated wavefunctions of the oscillator on a space with constant curvature $k$ are
\begin{equation}
 \Psi_{n_r,l,m}(r,\te,\phi;\k) = K_\k\, r^l(1 - \k\, r^2)^{\,(1/2\wt{\k})}
 {\cal P}_{n_r,l}(r;\k)Y_{lm}(\te,\phi)
\end{equation}
where ${\cal P}_{n_r,l}$ denotes the polynomial
$$
 {\cal P}_{n_r,l}(r;\k) = {}_2 F_1(-n_r,b_{n_r},c\,; \,\k\, r^2)
$$
with  $b_{n_r} = n_r + l + 1 +1/\wt{\k}$ (the value of $b_\k$ when
$a_\k=-n_r$), $c= l+3/2$ and $K_\k$ is a constant. These polynomials appear as a $k$-deformation of the $\k=0$ associated Laguerre polinomials.

  The set of wavefunctions $\Psi_{N_r,m}(r,\phi;\k)$ is a set of orthogonal functions with respect to the measure $d\mu_\k$ that is complete when $\k>0$; in the hyperbolic case, there is, in addition to the discrete spectrum,  also a continuous spectrum (the particular characteristics of the wavefunctions in the $\k<0$ case are discussed below).   
The constant $K_\k$ is obtained from the normalization
conditions which are given by
$$
 \int_{0}^{2\pi}\int_{0}^{\pi} \int_0^{r_\k} \bigl|\Psi_{n_r,l,m}\bigr|^2d\mu_\k
 = \int_{0}^{2\pi}\int_{0}^{\pi}\sin\te\,d\te\,d\phi\int_0^{r_\k} 
 \Bigl|\Psi_{n_r,l,m}\Bigr|^2 \Bigl(\frac{r^2}{\sqrt{1-\k\,r^2}}\Bigr)\,dr = 1
 \,,\quad  \k>0\,,
$$
and
$$
 \int_{0}^{2\pi}\int_{0}^{\pi}\int_{0}^{\infty}\bigl|\Psi_{n_r,l,m}\bigr|^2d\mu_\k
 = \int_{0}^{2\pi}\int_{0}^{\pi}\sin\te\,d\te\,d\phi\int_0^{\infty} 
 \Bigl|\Psi_{n_r,l,m}\Bigr|^2 \Bigl(\frac{r^2}{\sqrt{1-\k\,r^2}}\Bigr)\,dr = 1
 \,,\quad  \k<0\,,
$$
where we have used the notation  $r_\k=1/\sqrt{\k}$ in the $\k>0$
case.   We have obtained the following values for the radial integrals 
$$
  \int_0^{r_\k} r^{2 l}\,(1 - k\, r^2)^{\,(1/\k)} {\cal P}_{n_r,l}(r;\k)^2 
\Bigl(\frac{r^2}{\sqrt{1-\k\,r^2}}\Bigr)\,dr = K_{n_r}^+\,
\frac{\Gamma(l + 3/2) \,\Gamma(n_r + 1/2 + 1/\k) }{
\Gamma(n_r + l  + 1 + 1/\k)} 
 \,,\quad  \k>0\,,
$$
$$
  \int_0^{\infty} r^{2 l}\,(1 - k\, r^2)^{\,(1/\k)} {\cal P}_{n_r,l}(r;\k)^2 
\Bigl(\frac{r^2}{\sqrt{1-\k\,r^2}}\Bigr)\,dr = K_{n_r}^-\,
\frac{\Gamma(l + 3/2) \,\Gamma(1/|\k| -(2 n_r + 1 + l)) }{
\Gamma(1/|\k| + 1/2 - n_r )} 
 \,,\quad  \k<0\,,
$$
where $\Gamma(\cdot)$ denotes de Gamma function and the the coefficients $K_{n_r}^+$ and $K_{n_r}^-$ are given by 
$$
K_{n_r}^+ = \frac{\k^{-(1/2 + l)}  \,n_r !}{2\, (1 + \k\,(1 + l + 2 n_r) )(3/2 + l)_{n_r} }  \,,
$$
and
$$
  K_{n_r}^- = \frac{|\k|^{-(3/2 + l)}  \,n_r ! \,(1/|\k| - 2 n_r - l)_{n_r}}
  {2\, (3/2 + l)_{n_r} }  \,,
$$
where $(a)_{n_r}$ denotes the Pochhammer symbol $(a)_{n_r}=a\,(a+1)\dots (a+n_r-1)$. 

The  values of the energies are given by
\begin{equation}  
  \caE_n = (n + \frac{3}{2}) + \frac{1}{2}\,\wt{\k}\,n(n + 2)
  \,,{\quad} n = 2 n_r + l \,.
\end{equation}  
Two important properties are  (i) $\caE_n$ depends only on $n$, so the energy levels are degenerate with respect $n_r$ and $l$, and (ii) $\caE_n$ is the sum of the Euclidean value (correponding to $\k=0$) plus an additional term proportional to $n^2$ and with a coefficient depending directly of the curvature.

 In the Euclidean $\k=0$ case, the three-dimensional harmonic oscillator is just the sum of three independent one-dimensional oscillators; and the energy level $E_n = (n + 3/2)(\hbar\al)$ is  $(n+1)(n+2)/2$-fold degenerate since this is the number of ways that $n$ can be written as the sum of three non-negative integers (usually denoted by $n_x$, $n_y$, and $n_z$).  Now we have obtained that in the  non-Euclidean  $\k\ne 0$ case, the value of $n$, as a function of $n_r$ and $l$, is independent of $\k$; so the degeneracy of the energy levels is the same as in the Euclidean case (there exists accidental  degeneracy in addition to the essential degeneracy of a central potential).  Alternatively,  the value $(n+1)(n+2)/2$ can also be directly calculated by using the relations 
$$
 n = 2 n_r + l ,{\quad} n_r = 0,1,2,\dots,{\quad} l=0,1,2\dots,{\quad} m = -l, -l+1,\dots,l-1,l \,,    
$$
in a similar way as in the Euclidean case (see e.g.   \cite{CohenDiuLaloe} or \cite{BranJoa}). 

In the hyperbolic $\k<0$ case, as the radial integral is
defined on a infinite interval, the following property must be
satisfied
$$
 \lim_{r\to\infty}\, r \,\Bigl[\,r^l\,(1 - \k\,r^2)^{(1/2\,\k)}
 \,{}_2F_1(\k\,r^2)\,\Bigr]^2\,
 \Bigl(\frac{r^2}{\sqrt{1-\k\,r^2}}\Bigr) = 0  \,.
$$
The consequence is that if $\k<0$ then the the quantum numbers
$n_r$ and $l$ are limited by the condition
\begin{equation}  
 n = 2n_r + l < \frac{1}{\left|\wt{\k}\right|} - 1  \,,
\end{equation}  
and there are only $n_\k$ eigenvalues and eigenfunctions where
$n_\k$ denotes the greatest integer lower than $1/|\wt{\k}|-1$.

Figures 2 and 3 show the form of the radial functions $f(r,\k)\,(1
- \k\,r^2)^{1/(2\wt{\k})}$ for several values of $\k$ ($\k>0$ in figure
2 and $\k<0$ in figure 3).

The following two points summarize the main characteristics of the
energies of the bound states.

\begin{enumerate}

\item{} Spherical $\k>0$ case:

 The Hamiltonian $\widehat{H}(\k)$ describes a quantum oscillator
on the sphere $S_{\k}^3$ ($\k>0$). The oscillator possesses a
countable infinite set of bound states $\Psi_{n_r,l,m}(r,\te,\phi;\k)$,
with $n_r,l=0,1,2,\dots$, and the energy spectrum is unbounded, not
equidistant and with a gap between every two consecutive levels
that increases with $n$
\begin{eqnarray}
 &&\caE_0<\caE_1<\caE_2<\caE_3<\dots<\caE_n<\caE_{n+1}<\dots \cr
 &&\caE_{n+1} - \caE_n = 1 + \wt{\k}\,\bigl(n+\frac{3}{2}\bigr)\,.
\nonumber\end{eqnarray}
The oscillations of the wavefunctions are reinforced and the
values of the energies $E_n$ are higher than in the Euclidean
$\k=0$ case; i.e. $E_n(\k)>E_n(0)$.

\item{} Hyperbolic $\k<0$ case:

The Hamiltonian $\widehat{H}(\k)$ describes a quantum oscillator
on the hyperbolic space $H_{\k}^3$ ($\k<0$). The oscillator
possesses only a finite number of bound states
$\Psi_{n_r,l,m}(r,\te,\phi;\k)$, with $n=0,1,2,\dots, n_\k$,
$n_\k<1/|k|-1 $, and the energy spectrum is bounded, not
equidistant and with a gap between every two levels that decreases
with $n$
\begin{eqnarray*}
 &&\caE_0<\caE_1<\caE_2<\caE_3<\dots <\caE_{n_\k} \cr
 &&\caE_{n+1} - \caE_n = 1 - \left|\wt{\k}\right|\,\bigl(n+\frac{3}{2}\bigr)  \,.
\end{eqnarray*}
The oscillations of the wavefunctions are smoothed down and the
values of the energies $E_n$ are lower than in the Euclidean $\k=0$
case; i.e. $E_n(\k)<E_n(0)$.

In this $\k<0$ case there is also, in addition to the discrete (quantized) spectrum,   a continuous spectrum. Higher values  of the energy $\caE$ such that $\caE>\caE_{n_\k}$ correspond to scattering solutions.   These wavefunctions,  related to (nonpolinomial) hypergeometric functions,  are characterized by a continuous index (continuous value of the energy) and with orthogonalization relations given by the Dirac delta.  The total basis includes (as in a well with finite depth or in the Hydrogen atom) both functions labeled by a discrete index and functions labeled by a continuous index.  

\end{enumerate}

Figure 4 illustrates the two main characteristics of the energy
levels. The first one is that the values are higher in the
spherical case and lower in the hyperbolic plane, and the second
one is that the number of bound states is finite in the hyperbolic
case with the number increasing when $|\k|$ decreases. The plot
clearly shows that when the absolute value $|\k|$ decreases the
maximum of the curve moves into the upright, the number of bound
sates goes up and in the limit $\k\to 0$ the curve converges into
the straight line parallel to the diagonal (dashed line)
representing the Euclidean system.

The wavefunctions $\Psi_{N_r,m}(r,\te,\phi;\k)$ and the energies
$E_n$ show clear differences depending of the sign of $\k$ as it
was  expected. Nevertheless, if they are considered as
functions of the curvature $\k$ then all the changes are presented
in a smooth and continuous way.

\section{Final comments and outlook}

The Schr\"odinger equation is well defined for all the values of $\k$ but what introduce differences between the $\k>0$ and the $\k<0$ cases
is that in the spherical $S_\k^3$ ($\kappa>0$) case the space is compact 
and the oscillator possesses an infinite set of bound states; in the hyperbolic 
 $H_k^3$ ($\kappa<0$) case the potential $U_\k(r)$ is such that $U_\k(r)\,{\to}\,(1/2)(\al^2/|\k|)$ when $r\to\infty$ and the oscillator possesses only a finite number of bound states (for certain values of $|\k|$ only the fundamental level).

We finalize with two comments.

First, this paper is mainly concerned with the interface between 
geometry and quantum mechanics but it leads, in a natural way, to  
questions of functional analysis related to the theory of operators 
on Hilbert spaces.  In some respects these problems are similar to 
those studied in the standard Euclidean case, but depending on the 
sign and  the value of $\k$ these might go beyond and provide 
new aspects to the problem. The main point is the following: starting 
from the operator appearing in the (radial) Sturm-Liouville problem, 
which is symmetric in its natural domain, can it be extended to a 
self-adjoint operator in the $\k\ne 0$ case? Is this operator unique 
or  are there a family of extensions depending on parameters entering 
into the boundary conditions?  This is an open question in the 
$\k\ne 0$ case and we think that it deserves to be studied in more
detail.

Second, we point out once more that the existence of the harmonic
oscillator  is not a specific or special characteristic of the Euclidean
space but it is a well defined system in the three different spaces of
constant curvature. In fact, we have proved that, making use of
the curvature $\k$ as a parameter, there are not three different
harmonic oscillators but only one defined, at the same time, in the
three manifolds and endowed with properties depending smoothly
on the curvature.

\section*{Acknowledgments}

 JFC and MFR acknowledges support from from research projects MTM-2009-11154,  MTM-2010-12116-E (MEC, Madrid),  and DGA-E24/1 (DGA, Zaragoza), and MS from research projects MTM-2009-10751 (MEC, Madrid) and JCyL-GR224-08 (JCyL, Valladolid).   
The  authors are indebted to the referee for interesting remarks and helpful suggestions  which have definitely improved the presentation of this paper.

{\small

    }
\vfill\eject

\begin{figure}\centerline{
\includegraphics{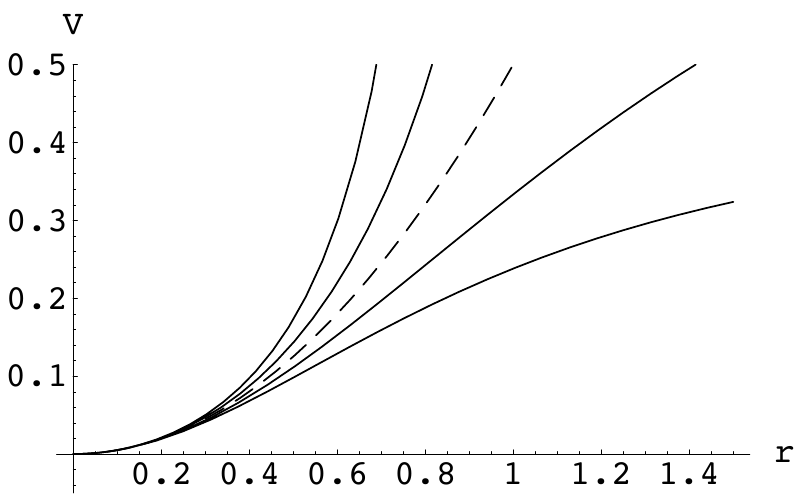}  }

\caption{Plot of the potential  $U_\k(r)$, $\al=1$, as a function of $r$, 
for $\k<0$ (lower curves), $\k=0$ (dash line), and $\k>0$ (upper curves).}
\label{Fig1}
\end{figure}

\begin{figure}\centerline{
\includegraphics{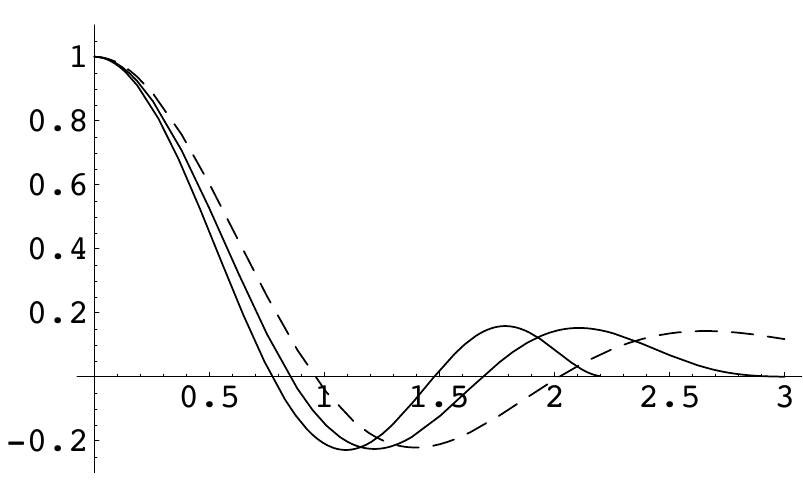}  }

\caption{Plot of three radial functions with quantum numbers $n_r=2$, $L=0$: 
the Euclidean function ${}_1F_1(\rho^2)\,\exp(-(1/2)\,\rho^2)$ ($\wt{\k}=0$, dashed curve)  and two $\k$-dependent functions ${}_2F_1(\wt{\k}\, \rho^2)(1 - \wt{\k}\,\rho^2)^{(1/2\,\wt{\k})}$ corresponding to $\wt{\k}=0.10$  and $\wt{\k}=0.20$.  
For very small values of the curvature $\wt{\k}$ the figure is very close to the $\wt{\k}=0$ radial curve but when the value of $\wt{\k}$ increases the oscillations
narrow and move into smaller values of $\rho$.}
\label{Fig2}
\end{figure}

\vfill\eject

\begin{figure}\centerline{
\includegraphics{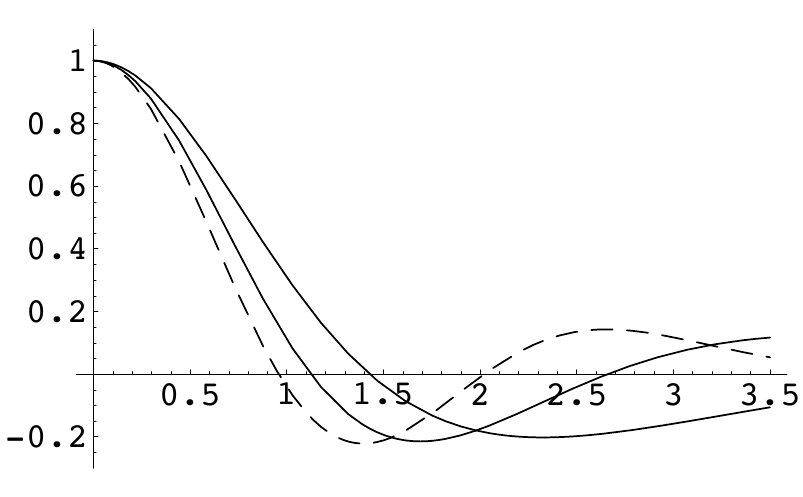}  }

\caption{Plot of three radial functions with quantum numbers $n_r=2$, $L=0$: the Euclidean function ${}_1F_1(\rho^2)\,\exp(-(1/2)\,\rho^2)$ ($\wt{\k}=0$, dashed curve) and two $\k$-dependent functions ${}_2F_1(\wt{\k}\, \rho^2)(1 - \wt{\k}\,\rho^2)^{(1/2\,\wt{\k})}$ corresponding to $\wt{\k}=-0.10$  and $\wt{\k}=-0.20$. 
When the absolute value $|\wt{\k}|$ increases then the oscillations soften and lengthen into greater values of $\rho$.} 
\label{Fig3}
\end{figure}

\begin{figure}\centerline{
\includegraphics{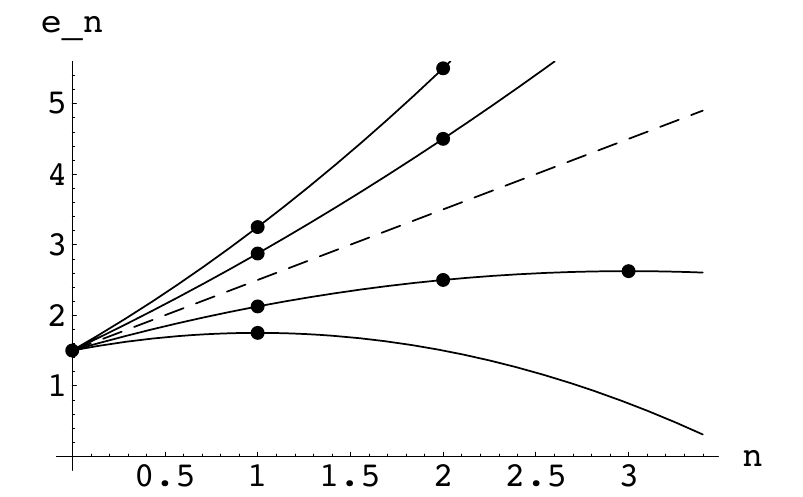}  }

\caption{ Plot of the energy $\caE_n$ as a function of $n$, $n = 2n_r + L$, for several values of the curvature with the thick points
$(n,\caE_n)$ representing the energies of the bound states. The upper
curves correspond to two spherical cases, $\wt{\k}=0.25$ and $\wt{\k}=0.50$;
the straight line parallel to the diagonal (dashed line)
represents the standard Euclidean case and the lower curves
represent two hyperbolical cases $\wt{\k}=-0.25$ (four bound levels)
and $\wt{\k}=-0.50$ (only two bound levels). }
\label{Fig4}
\end{figure}

\end{document}